\documentclass[graybox]{svmult}

\usepackage{mathptmx}       
\usepackage{helvet}         
\usepackage{courier}        
\usepackage{type1cm}        
\usepackage{makeidx}         
\usepackage{graphicx}        
\usepackage{multicol}        
\usepackage[bottom]{footmisc}
\usepackage{dcolumn}
\usepackage{bm}
\usepackage{graphicx}
\usepackage{amssymb}
\usepackage{amsmath}

\newcommand{\x}{{x}} 
\newcommand{\X}{{X}} 
\newcommand{\Y}{{X}} 
\newcommand{\G}{{Q}} 
\newcommand{\K}{{\rm K}} 
\newcommand{\Per}{{\cal T}} 

\newcommand{\sh}[5]{{\cal Y}^{#1#2#5}_{#3,#4}}
\newcommand{\Veig}{{\nu}} 
\newcommand{\V}{{\cal V}}

\begin{document}

\title*{Permutation-symmetric three-body O(6) hyperspherical harmonics in three
spatial dimensions}
\titlerunning{Permutation-symmetric three-body O(6) hyperspherical harmonics}
\author{Igor Salom and V. Dmitra\v sinovi\' c}
\institute{Igor Salom \at Institute of Physics, Belgrade
University, Pregrevica 118, Zemun, \\
P.O.Box 57, 11080 Beograd, Serbia, \email{isalom@ipb.ac.rs}
\and V. Dmitra\v sinovi\' c \at Institute of Physics, Belgrade
University, Pregrevica 118, Zemun, \\
P.O.Box 57, 11080 Beograd, Serbia,  \email{dmitrasin@ipb.ac.rs}}
\maketitle

\abstract{We have constructed the three-body permutation symmetric O(6) hyperspherical harmonics which can be used to solve the non-relativistic three-body Schr{\" o}dinger equation in three spatial dimensions.
We label the states with eigenvalues of the $U(1) \otimes SO(3)_{rot} \subset U(3) \subset O(6)$ chain
of algebras and we present the corresponding $K \leq 4$ harmonics.
Concrete transformation properties of the harmonics are discussed in some detail.
}
\section{Introduction}
\label{s:Introduction}

Hyperspherical harmonics are an important tool for dealing with quantum-mechanical three-body problem, being of a particular importance in the context of bound states \cite{Gronwall:1937,Bartlett:1937,DELVES:1958zz,Smith:1959zz,Simonov:1965ei,Iwai1987a}. However, before our recent progress \cite{Salom:2015}, a systematical construction of permutation-symmetric three-body hyperspherical harmonics was, to our knowledge, lacking (with only some particular cases being worked out -- e.g. those with total orbital angular momentum $L=0$, see Refs. \cite{Simonov:1965ei,Barnea:1990zz}).

In this note, we report the construction of permutation-symmetric three-body O(6) hyperspherical harmonics
using the $U(1) \otimes SO(3)_{rot} \subset U(3) \subset O(6)$ chain of algebras, where
$U(1)$ is the ``democracy transformation'',
or ``kinematic rotation'' group for three particles, $SO(3)_{rot}$ is the 3D rotation group,
and $U(3), O(6)$ are the usual Lie groups. This particular chain of algebras is mathematically very
natural, since the $U(1)$ group of
``democracy transformations'' is the only nontrivial (Lie) subgroup of full hyperspherical $SO(6)$
symmetry (the symmetry of nonrelativistic kinetic energy) that commutes with spatial rotations.
Historically, this chain was also suggested in
the recent review of the Russian school's work, Ref. \cite{Nikonov:2014}, and indicated by the
previous discovery of the dynamical $O(2)$ symmetry of the Y-string potential,
Ref. \cite{Dmitrasinovic:2009ma}. The name ``democracy transformations'' comes from the close
relation of these transformations with permutations: (cyclic) particle permutations form a discrete
subgroup of this $U(1)$  group.

The polynomial form of the obtained h.s.\ harmonics allows integrals of three (or more) harmonic
functions to be easily calculated. We demonstrate this in section \ref{sec:matrel}, where we give
tables of matrix elements of rotationally invariant permutation symmetric three particle harmonics
in the h.s.\ basis (these matrix elements are crucial for solving the Schr\"odinger equation in any
permutation symmetric potential).

\section{Three-body hyper-spherical coordinates}
\label{app:HS3D}

A natural set of coordinates for parametrization of three-body wave function $\Psi({\bm \rho},{\bm \lambda})$
(in the center-of-mass frame of reference) is given by the Euclidean relative position Jacobi vectors
${\bm \rho} = \frac{1}{\sqrt{2}}({\bf x}_{1} - {\bf x}_{2})$,
${\bm \lambda} = \frac{1}{\sqrt{6}}({\bf x}_{1} + {\bf x}_{2}- 2 {\bf x}_{3}).$
The overall six components of the two vectors can be seen as specifying a position in a six-dimensional configuration space $\x_{\mu} = ({\bm \lambda}, {\bm \rho})$, which, in turn, can be parameterized by hyper-spherical coordinates as $\Psi(R, \Omega_5)$. Here
$R = \sqrt{{\bm \rho}^2 + {\bm \lambda}^2}$ is the hyper-radius,
and five angles $\Omega_5$ parametrize a hyper-sphere in the
six-dimensional Euclidean space. Three ($\Phi_i;~ i=1,2,3$) of these five
angles ($\Omega_{5}$) are just
the Euler angles associated with the orientation in a three-dimensional
space of a spatial reference frame defined by the (plane of) three bodies; the remaining
two hyper-angles describe the shape of the triangle subtended
by three bodies; they are functions of three independent scalar three-body variables, e.g.
${\bm \rho} \cdot {\bm \lambda}$, ${\bm \rho}^2$, and ${\bm \lambda}^2$.
Due to the connection $R = \sqrt{{\bm \rho}^2 + {\bm \lambda}^2}$, this
shape-space is two-dimensional, and topologically equivalent to
the surface of a three-dimensional sphere. A spherical coordinate system can be further introduced in this shape space. Among various (in principle infinitely many) ways that this can be accomplished, the one due to Iwai \cite{Iwai1987a} stands out as the one that fully observes the permutation symmetry of the problem. Namely, of the two Iwai (hyper)spherical angles $(\alpha$, $\phi)$:
$(\sin \alpha)^2 = \,1 - \,\left(\frac{2 {\bm \rho} \times {\bm \lambda}}{R^{2}}\right)^{2}$,
$\tan \phi = \left(\frac{2 {\bm \rho} \cdot {\bm \lambda}}{{\bm \rho}^2
- {\bm \lambda}^2}\right)$, the angle $\alpha$ does not change under permutations, so that all permutation
properties are encoded in the $\phi$-dependence of the wave functions.

Nevertheless, in the construction of hyperspherical harmonics, we will, unlike the most of the previous attempts in this context, refrain from use of any explicit set of angles, and express harmonics as functions of cartesian Jacobi coordinates.

\section{O(6) symmetry of the hyper-spherical approach}
\label{s:kinHS}

The motivation for hyperspherical approach to the three-body problem comes from the fact that the equal-mass three-body kinetic energy $T$ is $O(6)$ invariant and
can be written as
\begin{eqnarray}
T_{\rm} &=& \frac{m}{2} \dot{R}^2 + \frac{K_{\mu\nu}^2}{2 m R^2}.
\label{e:kin1} \
\end{eqnarray}
Here, $K_{\mu\nu}$, ($\mu,\nu = 1,2,\ldots,6$) denotes the $SO(6)$ ``grand angular'' momentum tensor  
\begin{eqnarray}
K_{\mu\nu} &=& m \left({\bf x}_{\mu}{\dot{\bf x}}_{\nu} - {\bf x}_{\nu}{\dot{\bf x}}_{\mu}\right)
\nonumber \\
&=& \left({\bf x}_{\mu}{\bf p}_{\nu} - {\bf x}_{\nu}{\bf p}_{\mu}\right).
\label{e:K_tensor1} \
\end{eqnarray}
$K_{\mu\nu}$ has 15 linearly independent components, that contain, among themselves three components of the
``ordinary'' orbital angular momentum: ${\bf L} = {\bf l}_{\rho} + {\bf l}_{\lambda} = m
\left({\bm \rho} \times {\dot {\bm \rho}} + {\bm \lambda} \times {\dot {\bm \lambda}}\right)$.

It is due to this symmetry of the kinetic energy that the decomposition of the wave function and potential energy into $SO(6)$ hyper-spherical harmonics becomes a natural way to tackle the three-body quantum problem.

In this particular physical context, the six dimensional hyperspherical harmonics need to have some desirable  properties. Quite generally, apart from the hyperangular momentum K, which labels the O(6) irreducible representation,
all hyperspherical harmonics must carry additional labels specifying
the transformation properties of the harmonic with respect to
(w.r.t.) certain subgroups of the orthogonal group.
The symmetries of most three-body potentials, including the three-quark confinement ones, are:
parity, rotations and permutations (spatial exchange of particles).

Therefore, the goal is to find three-body hyperspherical harmonics with well defined transformation properties with respect to the symmetries.
Parity is directly related to $K$ value: $P = (-1)^{\rm K}$, the rotation symmetry implies that the hyperspherical harmonics must carry usual quantum numbers $L$ and $m$ corresponding to $SO(3)_{rot} \supset SO(2)$ subgroups and permutation properties turn out to be related with a continuous $U(1)$ subgroup of ``democracy transformations'', as will be discussed below.

\section{Labels of permutation-symmetric three-body hyper-spherical harmonics}
\label{s:HS}

We introduce the complex coordinates:
\begin{equation}
\X_i^\pm = \lambda_i \pm i \rho_i, \quad i =1,2,3.
\end{equation}
Nine of 15 hermitian $SO(6)$ generators
$K_{\mu\nu}$
in these new coordinates become
\begin{eqnarray}
i L_{ij} &\equiv&  
\X_i^+ \frac{\partial \ \  }{\partial \X_j^+} +
\X_i^- \frac{\partial \ \  }{\partial \X_j^-} - \X_j^+ \frac{\partial \ \  }{\partial \X_i^+} -
\X_j^- \frac{\partial \ \  }{\partial \X_i^-} 
, \label{opJinX}\\
2 Q_{ij} &\equiv& 
\X_i^+ \frac{\partial \ \  }{\partial \X_j^+} -
\X_i^- \frac{\partial \ \  }{\partial \X_j^-} + \X_j^+ \frac{\partial \ \  }{\partial \X_i^+} -
\X_j^- \frac{\partial \ \  }{\partial \X_i^-} 
. \label{opQinX}
\end{eqnarray}
Here $L_{ij}$ have physical interpretation of components of angular momentum vector ${\bf L}$. The symmetric tensor $Q_{ij}$
decomposes as (5) + (1) w.r.t. rotations, while the trace:
\begin{equation}
\G \equiv Q_{ii} = \sum_{i=1}^3 \X_i^+ \frac{\partial \ \  }{\partial \X_i^+}
- \sum_{i=1}^3\X_i^- \frac{\partial \ \  }{\partial \X_i^-}
\end{equation}
is the only scalar under rotations, among all of the $SO(6)$ generators. Therefore, the only mathematically justified choice is to take eigenvalues of this operator for an additional label of the hyperspherical harmonics. Besides, this trace $\G$ is the generator of the forementrioned democracy transformations, a special case of which are the cyclic permutations -- which in addition makes this choice particularly convenient on an route to construction of permutation-symmetric hyperspherical harmonics.
The remaining five components of the symmetric tensor $Q_{ij}$, together with three antisymmetric
tensors $L_{ij}$ generate the $SU(3)$ Lie algebra, which together with the single scalar $\G$ form
an $U(3)$ algebra, Ref. \cite{Nikonov:2014}.

Overall, labelling of the $O(6)$ hyper-spherical harmonics with labels $K, \G, L$ and $m$ 
corresponds to the subgroup chain $U(1) \otimes SO(3)_{rot} \subset U(3) \subset SO(6)$.
Yet, these four quantum numbers are in general insufficient to uniquely specify an $SO(6)$
hyper-spherical harmonic and an additional quantum number must be introduced to account for the remaining multiplicity. This is the multiplicity that necessarily occurs when $SU(3)$ unitary irreducible representations are labelled w.r.t.\ the chain $SO(2)\subset SO(3)\subset SU(3)$ (where $SO(3)$ is "matrix embedded" into $SU(3)$), and thus is well documented in the literature.
In this context 
the operator:
\begin{equation}
\V_{LQL} \equiv \sum_{ij} L_i \G_{ij} L_j
\label{e:JQJop}
\end{equation}
(where $L_i = \frac 12 \varepsilon_{ijk} L_{jk}$ and $\G_{ij}$ is given by
Eq. (\ref{opQinX})) has often been used 
to label the multiplicity of $SU(3)$ states. This operator commutes both with the
angular momentum $L_i$, and with the ``democracy rotation'' generator $Q$:
\[ \left[\V_{LQL} , L_i \right] = 0; ~~ \left[\V_{LQL} , Q \right] = 0\]
Therefore we demand that the hyper-spherical harmonics be eigenstates of this operator:
\[ \V_{LQL} \sh{K}{\G}{L}{m}{\Veig}
= \Veig \sh{K}{\G}{L}{m}{\Veig}
; ~~ \]
Thus, $\Veig$ will be the fifth label of the hyper-spherical harmonics, beside
the $(K, \G, L, m)$.


\section{Tables of hyper-spherical harmonics of given $K, \G, L, m$ and $\Veig$}
\label{a:tables}

Below we explicitly list all hyper-spherical harmonics for $\K \leq 4$, labelled by the quantum numbers $(K, \G, L, m, \Veig)$ (we will not delve here into lengthy details of the derivation of the expressions).
We list only the harmonics with $m=L$
and $Q\geq 0$, as the rest can
be easily obtained by acting on them with standard lowering operators
and by using the permutation symmetry properties of hyper-spherical harmonics:
$\sh{\K}{\G}{L}{m}{\Veig}(\lambda, \rho) =
(-1)^{\K-L} \sh{\K}{-\G}{L}{m}{-\Veig}(\lambda, -\rho)$.
We use the (more compact) spherical complex coordinates:
$\Y^{\pm}_0 \equiv \lambda_3 \pm i \rho_3$,
$\Y^{\pm}_{(\pm)} \equiv \lambda_1 \pm i \rho_1 + (\pm) (\lambda_2 \pm i \rho_2)$,
$|\Y^\pm|^2 =  \Y^\pm_+ \Y^\pm_- + (\Y^\pm_0)^2$, while we are also explicitly
writing out the $\K \leq 3$ harmonics in terms of Jacobi coordinates.

\small

\begin{equation*}
{\cal Y}_{0,0}^{0,0,0}(X)=\frac{1}{\pi ^{3/2}}
\end{equation*}

\begin{equation*}
{\cal Y}_{1,1}^{1,1,-1}(X)=\frac{\sqrt{\frac{3}{2}} {\Y}_+^+}{\pi ^{3/2} R}
=\frac{\sqrt{\frac{3}{2}} \left(\lambda_1 + i \left(\lambda_2 + \rho_1 +
i \rho_2\right)\right)}{\pi ^{3/2} \sqrt{\lambda_1^2 + \lambda_2^2
+ \lambda_3^2 + \rho_1^2 + \rho_2^2+\rho _3^2}}
\end{equation*}

\begin{equation*} {\cal Y}_{1,1}^{2,0,0}(X)
=\frac{\sqrt{3} \left({\Y}_+^- {\Y}_0^+ - {\Y}_+^+ {\Y}_0^-\right)}{\pi ^{3/2} R^2}
=\frac{2 \sqrt{3} \left(\lambda_3
\left(\rho_2 - i \rho _1\right) + i \left(\lambda_1+i \lambda _2\right)
\rho _3\right)}{\pi ^{3/2} \left(\lambda_1^2
+\lambda _2^2+\lambda _3^2+\rho _1^2+\rho _2^2+\rho _3^2\right)}
\end{equation*}

\begin{equation*} {\cal Y}_{2,2}^{2,0,0}(X)=\frac{\sqrt{3} {\Y}_+^+ {\Y}_+^-}{\pi ^{3/2} R^2}
=\frac{\sqrt{3}
\left(\lambda_1 + i \left(\lambda_2 + \rho_1 + i \rho_2\right)\right)
\left(\lambda_1 + i \lambda_2 - i \rho_1 + \rho_2\right)}{\pi^{3/2}
\left(\lambda_1^2 + \lambda_2^2 + \lambda_3^2 + \rho _1^2 + \rho_2^2 + \rho_3^2\right)}
\end{equation*}
\vspace{-0.7cm}

\begin{equation*} {\cal Y}_{0,0}^{2,2,0}(X)
= \frac{\sqrt{2} \left|{\Y}^+\right|^2}{\pi ^{3/2} R^2}
= \frac{\sqrt{2} \left(2 i \lambda_1 \rho_1 + 2 i \lambda_2 \rho_2 + 2 i \lambda_3 \rho_3
+ \lambda_1^2 +\lambda_2^2 + \lambda_3^2 - \rho_1^2 - \rho_2^2 - \rho_3^2\right)}{\pi^{3/2}
\left(\lambda_1^2 + \lambda_2^2 + \lambda_3^2 + \rho_1^2 + \rho_2^2 + \rho_3^2\right)}
\end{equation*}\vspace{-0.7cm}

\begin{equation*} {\cal Y}_{2,2}^{2,2,-3}(X)
= \frac{\sqrt{\frac{3}{2}} \left({\Y}_+^+\right)^2}{\pi ^{3/2} R^2}
= \frac{\sqrt{\frac{3}{2}} \left(\lambda_1 + i \left(\lambda_2 + \rho_1 +
i \rho_2\right)\right)^2}{\pi^{3/2}
\left(\lambda_1^2 + \lambda_2^2 + \lambda_3^2 + \rho_1^2 + \rho_2^2 + \rho _3^2\right)}
\end{equation*}
\vspace{-0.7cm}

\begin{eqnarray*}
{\cal Y}_{1,1}^{3,1,3}(X) &=&
\frac{\sqrt{6} \left({\Y}_+^- \left|{\Y}^+\right|^2
-\frac{1}{2} R^2 {\Y}_+^+\right)}{\pi ^{3/2} R^3} \\ 
&=&
\frac{\sqrt{6} \left(\left(\lambda_1 + i \lambda_2 - i \rho_1 + \rho_2\right)
\left(\left(\lambda_1 + i \rho_1 \right)^2
+ \left(\lambda_2 + i \rho _2 \right)^2 +
\left(\lambda_3 + i \rho_3 \right)^2 \right)
- \frac{1}{2} \left(\lambda_1 + i \left(\lambda_2 + \rho_1 + i \rho_2 \right)\right)
\right)}{\pi^{3/2}\left(\lambda_1^2 + \lambda_2^2 + \lambda_3^2 + \rho_1^2 + \rho_2^2 +
\rho_3^2\right)} \
\end{eqnarray*}
\vspace{-0.7cm}

\begin{equation*} {\cal Y}_{2,2}^{3,1,-5}(X)
= \frac{\sqrt{5}
{\Y}_+^+ \left({\Y}_+^- {\Y}_0^+-{\Y}_+^+ {\Y}_0^-\right)}{\pi ^{3/2} R^3}
=\frac{2 \sqrt{5} \left(\lambda _1+i \left(\lambda _2+\rho _1+i \rho _2\right)\right)
\left(\lambda _3 \left(\rho _2-i \rho _1\right)
+ i \left(\lambda _1+i \lambda _2\right) \rho _3\right)}{\pi ^{3/2}
\left(\lambda _1^2+\lambda _2^2+\lambda _3^2+\rho _1^2+\rho _2^2
+ \rho_3^2\right)^{3/2}}
\end{equation*}\vspace{-0.7cm}

\begin{equation*}
{\cal Y}_{3,3}^{3,1,-2}(X)
=\frac{\sqrt{15} \left({\Y}_+^+\right)^2 {\Y}_+^-}{2 \pi^{3/2} R^3}
=\frac{\sqrt{15} \left(\lambda _1+i \left(\lambda _2+\rho _1
+i \rho _2\right)\right)^2 \left(\lambda _1+i \lambda _2
-i \rho _1+\rho _2\right)}{2 \pi ^{3/2} \left(\lambda _1^2
+\lambda _2^2+\lambda _3^2+\rho _1^2+\rho _2^2+\rho _3^2\right)^{3/2}}
\end{equation*}\vspace{-0.7cm}

\begin{equation*} {\cal Y}_{1,1}^{3,3,-1}(X)
= \frac{\sqrt{3} {\Y}_+^+ \left|{\Y}^+\right|^2}{\pi ^{3/2} R^3}
= \frac{\sqrt{3} \left(\lambda _1+i \left(\lambda _2+\rho _1+i \rho _2\right)\right)
\left(2 i \lambda_1 \rho_1 +2 i \lambda_2 \rho_2 + 2 i \lambda_3 \rho_3 + \lambda_1^2
+ \lambda_2^2 + \lambda_3^2 - \rho_1^2 - \rho_2^2 - \rho_3^2\right)}
{\pi^{3/2} \left(\lambda_1^2 + \lambda_2^2 + \lambda_3^2 +
\rho_1^2 + \rho_2^2 + \rho_3^2\right)^{3/2}}
\end{equation*}\vspace{-0.7cm}

\begin{equation*} {\cal Y}_{3,3}^{3,3,-6}(X)
=\frac{\sqrt{5} \left({\Y}_+^+\right)^3}{2 \pi ^{3/2} R^3}
=\frac{\sqrt{5}
\left(\lambda _1+i \left(\lambda _2+\rho _1+i \rho _2\right)\right)^3}{2 \pi^{3/2}
\left(\lambda _1^2+\lambda_2^2 +\lambda _3^2+\rho _1^2+\rho _2^2+\rho _3^2\right)^{3/2}}
\end{equation*}

\begin{equation*} {\cal Y}_{0,0}^{4,0,0}(X)
=-\frac{\sqrt{3} \left(R^4-2 \left|{\Y}^-\right|^2
\left|{\Y}^+\right|^2\right)}{\pi ^{3/2} R^4}
\end{equation*}
\vspace{-0.7cm}

\begin{equation*} {\cal Y}_{2,2}^{4,0,-\sqrt{105}}(X)=\frac{-12 \sqrt{14} R^2 {\Y}_+^+ {\Y}_+^-
+\sqrt{105 \left(11-\sqrt{105}\right)} \left({\Y}_+^-\right)^2 \left|{\Y}^+\right|^2+\sqrt{105 \left(11+\sqrt{105}\right)}
\left({\Y}_+^+\right)^2 \left|{\Y}^-\right|^2}{14 \pi ^{3/2} R^4}
\end{equation*}\vspace{-0.7cm}

\begin{equation*} {\cal Y}_{2,2}^{4,0,\sqrt{105}}(X)=\frac{-12 \sqrt{14} R^2 {\Y}_+^+ {\Y}_+^-
+\sqrt{105 \left(11+\sqrt{105}\right)} \left({\Y}_+^-\right)^2 \left|{\Y}^+\right|^2+\sqrt{105
\left(11-\sqrt{105}\right)} \left({\Y}_+^+\right)^2 \left|{\Y}^-\right|^2}{14 \pi ^{3/2} R^4}
\end{equation*}\vspace{-0.7cm}

\begin{equation*} {\cal Y}_{3,3}^{4,0,0}(X)=
\frac{3 \sqrt{5} {\Y}_+^+ {\Y}_+^- \left({\Y}_+^- {\Y}_0^+-{\Y}_+^+ {\Y}_0^-\right)}{2 \pi ^{3/2} R^4}
\end{equation*}

\begin{equation*} {\cal Y}_{4,4}^{4,0,0}(X)=
\frac{3 \sqrt{\frac{5}{2}} \left({\Y}_+^+\right)^2 \left({\Y}_+^-\right)^2}{2 \pi ^{3/2} R^4}
\end{equation*}

\begin{equation*} {\cal Y}_{1,1}^{4,2,2}(X)
=\frac{3 \left({\Y}_+^- {\Y}_0^+-{\Y}_+^+ {\Y}_0^-\right) \left|{\Y}^+\right|^2}{\pi ^{3/2} R^4}
\end{equation*}

\begin{equation*} {\cal Y}_{2,2}^{4,2,2}(X)
=\frac{\sqrt{\frac{3}{7}} {\Y}_+^+ \left(5 {\Y}_+^- \left|{\Y}^+\right|^2-2 R^2 {\Y}_+^+\right)}{\pi ^{3/2} R^4}
\end{equation*}

\begin{equation*} {\cal Y}_{3,3}^{4,2,-13}(X)
=\frac{3 \sqrt{\frac{5}{2}} \left({\Y}_+^+\right)^2 \left({\Y}_+^-
{\Y}_0^+-{\Y}_+^+ {\Y}_0^-\right)}{2 \pi ^{3/2} R^4}
\end{equation*}

\begin{equation*} {\cal Y}_{4,4}^{4,2,-5}(X)
=\frac{\sqrt{15} \left({\Y}_+^+\right)^3 {\Y}_+^-}{2 \pi ^{3/2} R^4}
\end{equation*}

\begin{equation*}
{\cal Y}_{0,0}^{4,4,0}(X)
=\frac{\sqrt{3} \left|{\Y}^+\right|^4}{\pi ^{3/2} R^4}
\end{equation*}

\begin{equation*}
{\cal Y}_{2,2}^{4,4,-3}(X)
=\frac{3 \sqrt{\frac{5}{14}} \left({\Y}_+^+\right)^2 \left|{\Y}^+\right|^2}{\pi ^{3/2} R^4}
\end{equation*}

\begin{equation*} {\cal Y}_{4,4}^{4,4,-10}(X)
=\frac{\sqrt{15} \left({\Y}_+^+\right)^4}{4 \pi ^{3/2} R^4}
\end{equation*}

\normalsize

\section{Permutation symmetric hyper-spherical harmonics}
\label{s:permutation}

There is a small step remaining from obtaining the hyperspherical harmonics labelled by quantum numbers $(K, \G, L, m, \Veig)$ to achieving our goal, which is to construct hyperspherical functions with well-defined values of parity
$P
=(-1)^{\rm K}$, rotational group quantum numbers $(L, m)$,
and permutation symmetry $M$ (mixed), $S$ (symmetric), and $A$ (antisymmetric).\footnote{The mixed symmetry representation of the $S_3$ permutation group being two-dimensional,
there are two different state vectors (hyperspherical harmonics) in each mixed permutation
symmetry multiplet, usually denoted by $M_{\rho}$ and $M_{\lambda}$.} In this section we clarify how to obtain the latter as linear combinations of the former.

Properties under particle permutations of the functions $\sh{K}{\G}{J}{m}{\Veig}(\bm \lambda, \bm \rho)$
are inferred from the transformation properties of the coordinates $\X^\pm_i$:
under the transpositions (two-body permutations) $\{\Per_{12}, \Per_{23}, \Per_{31} \}$ of pairs
of particles (1,2),(2,3) and (3,1), the Jacobi coordinates transform as:
\begin{eqnarray}
\Per_{12}: & & {\bm \lambda} \rightarrow {\bm \lambda}, \quad  {\bm \rho} \rightarrow -
{\bm \rho},   \nonumber \\
\Per_{23}: & & {\bm \lambda} \rightarrow -\frac 12  {\bm \lambda} + \frac{\sqrt{3}}{2}
{\bm \rho}, \quad  {\bm \rho} \rightarrow \frac 12{\bm \rho} + \frac{\sqrt{3}}{2} {\bm \lambda},
\label{permutationsRL} \\
\Per_{31}: & & {\bm \lambda} \rightarrow -\frac 12  {\bm \lambda} - \frac{\sqrt{3}}{2}
{\bm \rho}, \quad  {\bm \rho} \rightarrow \frac 12{\bm \rho} - \frac{\sqrt{3}}{2}
{\bm \lambda}. \nonumber
\end{eqnarray}
That induces the following transformations of complex coordinates $\X^\pm_i$:
\begin{eqnarray}
\Per_{12}: & & \X^\pm_i \rightarrow \X^\mp_i,  \nonumber \\
\Per_{23}: & & \X^\pm_i \rightarrow e^{\pm \frac{2 i \pi}{3}}\X^\mp_i, \label{PermutationsX} \\
\Per_{31}: & & \X^\pm_i \rightarrow e^{\mp \frac{2 i \pi}{3}}\X^\mp_i. \nonumber \
\end{eqnarray}

None of the quantum numbers $\K, L$ and $m$ change under permutations of particles, whereas the values
of the ``democracy label'' $\G$ and multiplicity label $\Veig$ are inverted under all transpositions:
$\G \rightarrow -\G, \Veig \rightarrow -\Veig$.

Apart from the changes in labels, transpositions of two particles generally also result
in the appearance of an additional phase factor multiplying the hyper-spherical harmonic.
For values of $\K, \G, L$ and $m$ with no multiplicity,
we readily derive (Ref.\ \cite{Salom:2015}) the following transformation properties of h.s.
harmonics under (two-particle) particle transpositions:
\begin{eqnarray} \Per_{12}: & &
\sh{K}{\G}{L}{m}{\Veig}
\rightarrow
(-1)^{K-J}\sh{K}{,-\G}{L}{m}{,-\Veig},
\nonumber \\
\Per_{23}: & & \sh{K}{\G}{L}{m}{\Veig}
\rightarrow (-1)^{K-L} e^{\frac{2 \G i \pi}{3}}\sh{K}{,-\G}{L}{m}{,-\Veig}, 
\label{PermutationsSH}  \\
\Per_{31}: & & \sh{K}{\G}{L}{m}{\Veig}
\rightarrow (-1)^{K-L}
e^{- \frac{2 \G i \pi}{3}}\sh{K}{,-\G}{L}{m}{,-\Veig}. 
\nonumber
\end{eqnarray}

There are three distinct irreducible representations of the $S_3$
permutation group - two one-dimensional (the symmetric S and the antisymmetric A ones)
and a two-dimensional (the mixed M one).
In order to determine to which representation of the permutation group
any particular h.s.\ harmonic $\sh{K}{\G}{L}{m}{\Veig}$ 
belongs, one
has to consider various cases, with and without multiplicity, see Ref.\ \cite{Salom:2015};
here we simply state the results of the analysis conducted therein. 
The following linear combinations of the h.s.\ harmonics,
\begin{equation} \sh{K}{|\G|}{L}{m, \pm}{\Veig}
\equiv \frac{1}{\sqrt{2}}\left(\sh{K}{|\G|}{L}{m}{\Veig}
\pm (-1)^{K - L}\sh{K}{,-|\G|}{L}{m}{,-\Veig}
\right).
\end{equation}
are no longer eigenfunctions of operators $\G$ and $\V_{LQL}$ but are (pure sign) eigenfunctions of
the transposition $\Per_{12}$ instead:
\[ \Per_{12}:
\sh{K}{|\G|}{L}{m, \pm}{\Veig}
\to \pm \sh{K}{|\G|}{L}{m, \pm}{\Veig}. \] 
They are the appropriate h.s.\ harmonics with well-defined permutation properties:
\begin{enumerate}
\item for $\G \not\equiv 0$ (mod 3), the harmonics $\sh{K}{|\G|}{L}{m, \pm}{\Veig}$
belong to the mixed representation M, where the $\pm$ sign determines the $M_{\rho}, M_{\lambda}$
component,
\item for $\G \equiv 0$ (mod 3), the harmonic $\sh{K}{|\G|}{L}{m, +}{\Veig}$
belongs to the symmetric representation S and $\sh{K}{|\G|}{L}{m, -}{\Veig}$
belongs to the antisymmetric representation A.
\end{enumerate}
The above rules define the representation of $S_3$ for any given h.s. harmonic. 

\vspace{-0.5cm}

\section{Matrix elements}
\label{sec:matrel}

Decomposition into hyperspherical harmonics with manifest permutation properties
highly simplifies solving of Schr\"odinger's equation. The benefits are most notable
when the three-body potential is permutation symmetric.
The decomposition of any such potential into h.s.\ harmonics has low number of nonzero
components due to the permutation symmetry constraints:
e.g.\ up to $K \leq 11$ the only h.s.\ harmonics that can appear in such decomposition
are $\sh{0,}{|0|,}{0}{0, +}{0} $, $\sh{4,}{|0|,}{0}{0, +}{0}$, $\sh{8,}{|0|,}{0}{0, +}{0}$
and $\sh{6,}{|6|,}{0}{0, +}{0}$. In tables \ref{tab:Kleq4diag} and \ref{tab:drcc44b}, we show the nonzero matrix elements between states of same $K$, $K \leq 4$.
These matrix elements are sufficient to evaluate the matrix elements of permutation symmetric
sums of arbitrary one-, two- and three-body operators, such as the three-body potential,
and thus to solve Schr\"odinger's equation in the first order
of perturbation theory (in $K \leq 4$ subspace).

These harmonics have been applied in Ref. \cite{Prague:2015} to three homogeneous potentials.
The ordering of eigenstates (``pattern'' of the spectrum) depends on the O(6) symmetry-breaking,
which in turn is determined by the hyperspherical expansion coefficients of the three-body potential.
These coefficients depend on the dynamical ``remnant'' symmetries of the potential.
Thus, for example the so-called Y-string potential has an $O(2)$ dynamical symmetry,
Ref. \cite{Dmitrasinovic:2009ma}, that is absent in potentials that are pairwise sums of
single-power-law terms (for powers different than the second one).


\begin{table}[tbh]
\begin{center}
\caption{The values of
the three-body potential hyper-angular diagonal matrix elements
$\langle \, \sh{4,}{|0|,}{0}{0, +}{0} \,\rangle_{\rm ang}$, $\langle \, \sh{6,}{|6|,}{0}{0, +}{0} \,\rangle_{\rm ang}$ and $\langle \, \sh{8,}{|0|,}{0}{0, +}{0} \,\rangle_{\rm ang}$,
for $K \leq 4$ states (for all allowed orbital waves L). The correspondence
between the $S_3$ permutation group irreps. and SU(6)$_{FS}$ symmetry multiplets
of the three-quark system: $S \leftrightarrow 56$, $A \leftrightarrow 20$
and $M \leftrightarrow 70$. The table values are independent of the angular moment projection $m$.}
\begin{tabular}{cccccc}
\hline \hline K
& $\left(K, |Q|, L, \nu, \pm \right)$ & $[SU(6),L^P]$ & $\pi \sqrt{\pi} \langle \, \sh{4,}{|0|,}{0}{0, +}{0} \,\rangle_{\rm ang}$ & $\pi \sqrt{2\pi} \langle \sh{6,}{|6|,}{0}{0, +}{0}\,\rangle_{\rm ang}$ & $\pi \sqrt{\pi} \langle \, \sh{8,}{|0|,}{0}{0, +}{0} \,\rangle_{\rm ang}$ \\
\hline
\hline
2 & $\left(2, |2|, 0, ~~0, \pm\right)$ & $[70,0^+]$ & $\frac{1}{\sqrt{3}}$ & 0 & 0\\ 
2 & $\left(2, |0|, 2, ~~0, +\right)$ & $[56,2^+]$ & $\frac{\sqrt{3}}{5}$  & 0 & 0\\ 
2 & $\left(2, |2|, 2, - 3, \pm\right)$ & $[70,2^+]$ & $-\frac{1}{5\sqrt{3}}$  & 0 & 0\\ 
2 & $\left(2, |0|, 1, ~~0, -\right)$ & $[20,1^+]$ & $-\frac{1}{\sqrt{3}}$  & 0 & 0\\ 
3 & $\left(3, |3|, ~1, -1, -\right)$ & $[20,1^-]$ & $~\frac{1}{\sqrt{3}}$ & $-1$  & 0\\ 
3 & $\left(3, |3|, ~1, -1, +\right)$ & $[56,1^-]$ & $~\frac{1}{\sqrt{3}}$ & $~~1$  & 0\\ 
3 & $\left(3, |1|, ~1, ~~~3, \pm\right)$ & $[70,1^-]$ & $~0$ & $~~0$  & 0\\ 
3 & $\left(3, |1|, ~2, -5, \pm\right)$ &  $[70,2^-]$ & $-\frac{1}{\sqrt{3}}$ & $~~0$  & 0\\ 
3 & $\left(3, |1|, ~3, -2, \pm\right)$ & $[70,3^-]$ & $~\frac{5}{7 \sqrt{3}}$ & $~~0$  & 0\\ 
3 & $\left(3, |3|, ~3, -6, +\right)$ & $[56,3^-]$ & $-\frac{\sqrt{3}}{7}$ & $~~\frac{2}{7}$ & 0\\ 
3 & $\left(3, |3|, ~3, -6, -\right)$ & $[20,3^-]$ & $-\frac{\sqrt{3}}{7}$ & $-\frac{2}{7}$ & 0\\ 
4 & $\left(4, |4|, 0, ~~0, \pm \right)$ & $[70,0^+]$ & $~\frac{\sqrt{3}}{2}$ & 0 & $~\frac{1}{2\sqrt{5}}$ \\
4 & $\left(4, |0|, 0, ~~0, + \right)$ & $[56,0^+]$ & $~0$ & 0 & $~\frac{2}{\sqrt{5}}$ \\
4 & $\left(4, |2|, 1, 2, \pm \right)$ & $[70,1^+]$ & $~~0$ & 0 & $-\frac{1}{\sqrt{5}}$ \\
4 & $\left(4, |0|, 2, \sqrt{105}, +\right)$ & $[56,2^+]$ & $-\frac{12 \sqrt{3}}{35}$ & 0 & $\frac{\sqrt{5}}{7}$ \\
4 & $\left(4, |0|, 2, \sqrt{105}, -\right)$ & $[20,2^+]$ & $0$ & 0 & $-\frac{1}{\sqrt{5}}$ \\
4 & $\left(4, |2|, 2, 2, \pm \right)$ & $[70,2^+]$ & $\frac{4 \sqrt{3}}{35}$ & 0 & $\frac{\sqrt{5}}{7}$ \\
4 & $\left(4, |4|, 2, -3, \pm \right)$ & $[70^{'},2^+]$ & $~\frac{2 \sqrt{3}}{7}$ & 0 & $-\frac{1}{7 \sqrt{5}}$ \\
4 & $\left(4, |2|, 3, -13, \pm \right)$ & $[70,3^+]$ & $-\frac{5 \sqrt{3}}{14}$ & 0 & $~\frac{1}{14 \sqrt{5}}$ \\
4 & $\left(4, |0|, 3, ~~0,- \right)$ & $[20,3^+]$ & $-\frac{3 \sqrt{3}}{14}$ & 0 & $-\frac{\sqrt{5}}{14}$ \\
4 & $\left(4, |0|, 4, ~~0, + \right)$ & $[56,4^+]$ & $\frac{5 \sqrt{3}}{14}$ & 0 & $~\frac{3}{14 \sqrt{5}}$ \\
4 & $\left(4, |2|, 4, -5, \pm \right)$ & $[70,4^+]$ & $\frac{3 \sqrt{3}}{14}$ & 0 & $-\frac{\sqrt{5}}{42}$ \\
4 & $\left(4, |4|, 4, -10, \pm \right)$ & $[70^{'},4^+]$ & $-\frac{3 \sqrt{3}}{14}$ & 0  & $\frac{1}{42 \sqrt{5}}$ \\
\hline
\end{tabular}
\label{tab:Kleq4diag}
\end{center}
\end{table}
\vspace{-0.5cm}
\begin{table}[tbh]
\begin{center}
\caption{The values of the off-diagonal matrix elements of the hyper-angular part of the
three-body potential
$\pi \sqrt{2\pi} \langle [SU(6)_{f},L_{f}^P] |\, \sh{6,}{|6|,}{0}{0, +}{0}
\,| [SU(6)_{i},L_{i}^P] \rangle_{\rm ang}$,
for various K = 4 states (for all allowed orbital waves L).}
\begin{tabular}{cccc}
\hline \hline K & $[SU(6)_{f},L_{f}^P]$
& $[SU(6)_{i},L_{i}^P]$ & $\pi \sqrt{2\pi} \langle  \sh{6,}{|6|,}{0}{0, +}{0}  \rangle_{\rm ang}$ \\
\hline
\hline
4 & $[70,2^+]$ & $[70^{'},2^+]$ & $\frac{6}{7} \sqrt{\frac{6}{5}}$ \\ 
4 & $[70^{'},2^+]$ & $[70,2^+]$ & $\frac{6}{7} \sqrt{\frac{6}{5}}$ \\ 
4 & $[70,4^+]$ & $[70^{'},4^+]$ & $\frac{8}{21}$ \\ 
4 & $[70^{'},4^+]$ & $[70,4^+]$ & $\frac{8}{21}$ \\ 
\hline
4 & $[20,L^+]$ & $[20,L^+]$ & $~~0$ \\ 
4 & $[56,L^+]$ & $[56,L^+]$ & $~~0$ \\ 
4 & $[20,L^+]$ & $[56,L^+]$ & $~~0$ \\ 
\hline
\end{tabular}
\label{tab:drcc44b}
\end{center}
\end{table}

\section{Summary and Discussion}
\label{s:Summary}

In this paper we have reported on our recent construction of permutation
symmetric three-body $SO(6)$ hyper-spherical harmonics. In the section \ref{a:tables}
we have displayed explicit forms the harmonic functions labelled by quantum numbers
$K, \G, L, m$ and $\Veig$, postponing explanation of their derivation to \cite{Salom:2015}.
In section \ref{s:permutation} we demonstrated that simple linear combinations
$\sh{K}{|\G|}{L}{m, \pm}{\Veig}$ of these functions have well defined permutation properties
and, then in section \ref{sec:matrel} we calculated their (non-vanishing) matrix elements
that are sufficient for the evaluation of two- and three-body operators' matrix elements.
Our $O(6)$ permutation-symmetric three-body hyperspherical harmonics appear to be the first of
their kind in the literature.
Symmetrized three-body hyper-spherical harmonics have been pursued before, albeit without emphasis
on the the ``kinematic rotation'' $O(2)$ symmetry label $\Veig$.

Finally, a word about other attempts at symmetrized hyperspherical harmonics is due.
To our knowledge, aside from the special case $L=0$ results of Simonov, Ref. \cite{Simonov:1965ei}
and $L=1$ of Barnea and Mandelzweig, Ref. \cite{Barnea:1990zz}, several other attempts,
based on so-called ``tree pruning'' techniques, exist in the literature,
Refs. \cite{delAguila:1980,Aquilanti:1986,Kuppermann2005} (see also the extensive reference list
and commentary 
in the review Ref. \cite{Nikonov:2014}), beside the recursively symmetrized
N-body (with N$>$3) hyperspherical harmonics of Barnea and Novoselsky, Refs. \cite{Barnea1997,Barnea1998}.
The latter are based on the  $O(3) \otimes S_N  \subset O(3N-3)$ chain of algebras, which does
not explicitly include the ``kinematic rotation''/``democracy'' $O(2)$ symmetry.
In Ref. \cite{Barnea1998} only the two-body operator matrix elements were evaluated,
however, and none of three-body ones. 

Whereas, in the final instance, Barnea and Novoselsky's work must be
related to ours, we note the following basic differences:
a) their work is based on a different chain of algebras: $O(3) \otimes S_3  \subset O(6)$
vs. our $U(1) \otimes SO(3)_{rot} \subset U(3) \subset SO(6)$;
b) theirs is an essentially recursive-numerical method relying on the knowledge of tables of
the symmetric group $S_3$ Clebsch-Gordan coefficients, whereas ours is a
recursive-algebraic/group-theoretical approach;
c) their $S_{3}$ hyperspherical states are expressed in terms of $S_{2}$ hyperspherical states,
that are coupled, via the ``tree'' method;
whereas ours makes no reference to any two-body substate;
d) they evaluated only two-body matrix elements, whereas our method allows us to evaluate
matrix elements of all, including 
three-body operators, such as the area (of the triangle).

%
\begin{acknowledgement}
This work was financed by the Serbian Ministry of Science and
Technological Development under grant numbers OI 171031, OI 171037 and III 41011.
\end{acknowledgement}
\end{document}